# AN APPROXIMATE CALCULATION OF ACTIVE PIXEL SENSORS TEMPORAL NOISE FOR HIGH ENERGY PHYSICS

**Nicolas T. Fourches**
DSM/DAPNIA/SEI , CEA Saclay , Bat 141

## A – Schematic of the frontend

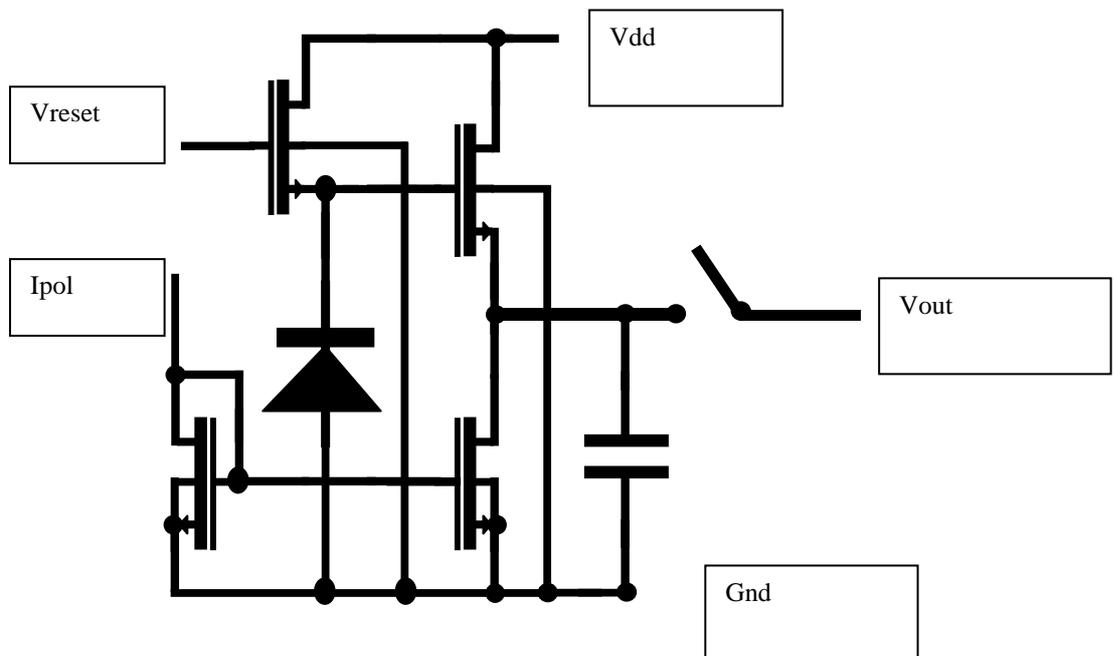

For the frontend temporal noise analysis we used this schematic because it corresponds to the operation of the real circuit when the noise is solely considered . The switch is present to take into account the fact that the noise is only observed during the time T . (The switch is then closed). In practical it is the signal that is output on the capacitance during T. The signal passes in an nmos before it reaches the capacitance . While the switch is closed the practical circuit is equivalent to this circuit , for the noise determination .

## B- Noise source in the frontend

Basically we have considered two principal source of noise in the design:
- The noise from the detector (represented here by a diode ), it is a shot noise if the detector is a good pin junction, it can be a resistor thermal noise if the characteristic is similar to that of a resistor.

- The noise from the transistors, from a simple analysis of the frontend, the contribution of the channel thermal noise of the source follower transistor should be dominant. The reset transistor has an insignificant contribution during measurement as it is switched off. It is possible to reduce significantly the contribution of the current mirror, the output of the mirror behaving as a nearly perfect current source with a transistor adequate geometry. The 1/f noise from the follower should be taken into account in a further step.

### C – Noise calculations : methods and results

Noise calculations should take into account the time dependent filtering due to the observation of noise during a limited time T. This duration corresponds to the time the output nmos switch is on in the real design. To overcome this problem one can use the Campbell theorem. Only the direct calculation of the series noise (thermal noise from the follower transistor) without filtering is possible from the noise spectral densities. The so called parallel noise from the detector (current noise) cannot be derived directly from the spectral densities. We made the calculation of the noise from the follower and the detector to determine the rms voltage noise at the capacitance node. The value of the noise voltage power due to the channel thermal noise is given by :

$$<V_{out}^2> = 2k\theta g_m \left(1+\frac{C_{gd}}{C_{gs}}\right)^2 \frac{1}{B^2} \int_0^T \exp\left(-\frac{2g_m C_{gd}}{BC_{gs}}t\right) dt$$

where $\theta$ is the absolute temperature:

where B is given by :

$$B = C + C_{gd} + \frac{C_{gd}}{C_{gs}} C$$

This results in :

$$<V_{out}^2> = k\theta \left(1+\frac{C_{gd}}{C_{gs}}\right)^2 \frac{C_{gs}}{BC_{gd}} \left(1-\exp\left(-\frac{2g_m C_{gd}}{BC_{gs}}T\right)\right)$$

Here gm is the transconductance of the NMOS follower, C is the output capacitance, Cgd, Cgs are the gate drain capacitance, the gate source capacitance respectively.

As we can see the noise power increases with T but tends to a limit of the form : kT/C The simulations will show that the response of the frontend to a charge pulse of the form Qδ(t) is a transient which settles after a time duration. If this time duration is lower than T we may determine the noise equivalent charge that is given by :

$$ENC_{series}^2 = k\theta \left(1+\frac{C_{gd}}{C_{gs}}\right)^2 \frac{C_{gs}C_{gd}}{B} \left(1-\exp\left(-2T\frac{g_m}{B}\frac{C_{gd}}{C_{gs}}\right)\right)$$

For the calculation of the noise power we determined the transfer function that is of the form:

$$\frac{v_{out}}{i_c}(s) = \frac{\left(1 + \frac{C_{gd}}{C_{gs}}\right)}{Bs + \frac{C_{gd}}{C_{gs}} g_m}$$

Where $i_c$ is the channel current.

For the impulse response of the frontend circuit the transfer function was calculated, we obtain:

$$\frac{v_{out}}{i_n}(s) = \frac{1}{C_{gd} s} + \frac{\beta}{\left(1 + s\left(\frac{A}{g_m C_{gd}}\right)\right) g_m C_{gd}}$$

$$\frac{v_{out}}{i_n}(s) = \frac{C_{gd}}{s} + \frac{\beta}{\left(1 + s\left(\frac{A}{g_m C_{gd}}\right)\right) g_m C_{gd}}$$

The time constant appears in this expression it is equal to:

$$A/(g_m C_{gd})$$

Where A and β are given by:

$$A = C C_{gs} + C C_{gd} + C_{gd} C_{gs}$$

$$\beta = C_{gs} - \frac{A}{C_{gd}}$$

The noise power results from the summation of three terms, we have calculated the first term and the summation of the last two, each corresponding to the contribution of the noise from the detector (parallel noise), we obtain:

$$<V_{out1//}^2> = \frac{2 q I_{leak} T}{C_{gd}^2}$$

$$<V_{out3//}^2> = q I_{leak} (1 - \exp(-2T g_m \frac{C_{gd}}{A})) \frac{1}{A g_m C_{gd}} (C_{gs} - \frac{A}{C_{gd}})^2$$

$$<V_{out2//}^2> = 4 q I_{leak} (\exp(-T g_m \frac{C_{gd}}{A}) - 1) \frac{1}{g_m C_{gd}^2} (C_{gs} - \frac{A}{C_{gd}})$$

$$<V^2_{out\,2+3//}> = 2q\,I_{leak}(1-\exp(-2T g_m \frac{C_{gd}}{A}))\frac{\beta}{g_m C_{gd}}\left(\frac{\beta C_{gd} - 2A}{2A C_{gd}}\right)$$

The equivalent noise charge can be determined in the same way as for the series noise: For example the first term is :

$$ENC_{1//}^2 = 2q\,I_{leak}\,T$$

the same calculation applies for the two other terms .
The total ENC squared is the sum of these three terms . In practical simple hand calculations show that only the first term is dominant so we can consider $2qI_{leak}T$ as being the ENC squared contribution from the detector part.

### D- Computation of noise numerical values

We compute here the values of the noise with reasonable assumptions concerning the capacitance values, the transconductance and the value of the leakage current and observation time T. Let us have:
$C_{gd} = C_{gs} = 20$ fF , this is reasonable for a 10/1 (µm/µm) dimensioned transistor
C= 1pF which is a value often used in APS.
We take $g_m=100\mu A/V$ as a reasonable value for the transconductance, 1ms as the value for the noise observation time, this is clearly an overestimation, and 10 fA =$I_{leak}$ for the leakage current through the detector (diode) and the reset transistor, which is a reasonably high assumption.

For the series noise we obtain:

ENCs = 14 e-

For the parallel noise we obtain:

ENC// = 11 e-

This is if the observation time is greater than the settling time of the signal which is the case with 1ms or even less.
The series voltage noise is then equal to:

$(<V_{out}^2>)^{0.5}$ = 110 µV rms

The parallel noise voltage is below:

$(<V_{out}^2>)^{0.5}$ = 88.5 µV rms

The contribution of the noise from the detector is lower than the contribution from the transistors, even with the worst case conditions that have been taken.

### E- Conclusions

To conclude this study a comparison can be made with the charge created by a minimal ionizing particle (MIP), approximately 83 e- per micrometer are deposited by these particles. For a 25 micrometers thickness the charge generated is of the order of 2000 e- which gives a reasonably high signal to noise ratio (2000/18) is approximately 112.

Note added: 27/06/2006.

Since these calculations were issued in 2000 measurements were made on MAPS. On neutron irradiated MAPS the leakage current of the pixels were of the order of 5 fA, this corresponds to around 8 e- of parallel equivalent noise charge. The estimation of the total input referred noise is ENC= 16 e- for a neutron irradiated chip, a very slightly higher value than that measured on the un-irradiated chips (~12 e-).